\documentclass[aps,prd,preprintnumbers,nofootinbib,twocolumn]{revtex4}
\usepackage{bm}
\usepackage{latexsym}
\usepackage{dcolumn}
\usepackage{amsmath,amsfonts,amssymb}
\usepackage{graphicx,epsfig}
\usepackage{amsthm}

\def\be {\begin{equation}}
\def\ee {\end{equation}}
\def\bea {\begin{eqnarray}}
\def\eea {\end{eqnarray}}
\def\bc {\begin{center}}
\def\ec {\end{center}}
\def\bfg {\begin{figure}}
\def\efg {\end{figure}}
\def\bi {\begin{itemize}}
\def\ei {\end{itemize}}

\def\la {\label}
\def\le {\left}
\def\ri {\right}

\def\vs {\vspace}

\def\o  {\omega}

\def\beq{\begin{equation}}
\def\eeq{\end{equation}}
\def\br{\begin{eqnarray}}
\def\er{\end{eqnarray}}
\newcommand{\eel}[1] {\label{#1}\end{equation}}

\newcommand{\bdm}{\begin{displaymath}}
\newcommand{\edm}{\end{displaymath}}

\begin{document}
\title{Can MOND type hypotheses be tested in a free fall laboratory environment?}

\author{Saurya Das\footnote{Also a member of the Theoretical Physics Group.
}} \email[email: ]{saurya.das@uleth.ca}
\author{S.N. Patitsas} \email[email: ]{steve.patitsas@uleth.ca}

\affiliation{Department of Physics and Astronomy,
University of Lethbridge, 4401 University Drive,
Lethbridge, Alberta, Canada T1K 3M4 \\}

\begin{abstract}
The extremely small accelerations of objects required for the
the onset of modified Newtonian dynamics, or MOND, makes
testing the hypothesis in conventional
terrestrial laboratories virtually impossible. This is
due to the large background acceleration of Earth, which is transmitted to the
acceleration of test objects within an apparatus. We show however, that it may
be possible to test MOND-type hypotheses with experiments using a
conventional apparatus capable of tracking very small accelerations of its
components, but performed in locally inertial frames such as artificial satellites and other
freely falling laboratories. For example, experiments involving an optical
interferometer or a torsion balance in these laboratories would show
nonlinear dynamics, and displacement amplitudes larger than expected.
These experiments may also be able to test potential violations of the
strong equivalence principle by MOND and to distinguish between its two possible interpretations
(modified inertia and modified gravity).
\end{abstract}

\maketitle



\null
The asymptotic flatness of galaxy rotation curves in spiral galaxies and the
related apparent mass discrepancy according to Newtonian dynamics
led to the proposal of dark matter \cite{zwicky}. Dark matter has also been inferred from
observations of apparent magnitudes of Type Ia supernovae at large redshifts within the
standard Freedman-Robertson-Walker cosmological paradigm and could account for as much as about
28\% of the mass-energy density of the observable Universe \cite{perlmutter}. Although
several dark matter candidates have been proposed, including weakly interacting massive
particles, axions etc., none has been directly observed so far.

Another proposal known as modified Newtonian dynamics (MOND) postulates the modification of
the law of gravity to \begin{equation}
a~\mu\le(a/a_0\ri) = \frac{GM}{r^2}~, \la{mond1}
\end{equation}
where $a$ is the acceleration of an object at a distance $r$ from a
gravitating mass $M$; the function $\mu$ is such that
$\mu(a/a_0)=1$ when $a\gg a_0$ (Newtonian regime) and
$=a/a_0$ when $a\leq a_0$ (deep MOND regime); and
$a_0\approx 1.2\times 10^{-10}~m/s^2$
is a characteristic, acceleration parameter separating the two regimes~\cite{milgrom1}.
If the acceleration of a system $a$ is written as $a=10^n a_0$,
then we define $n \geq 1$ as the Newtonian regime;
$n \approx 0$ as the onset of MOND\footnote{
An often used function is $\mu(x)=x/\sqrt{1+x^2}$~\cite{scarpa}.}; and $n\leq 0$ as the fully, or deep, MOND regime.
For circular motion, it follows from Eq.~(\ref{mond1})
that in the deep MOND regime, $v^4 = a_0 GM$ which is a constant and resembles the baryonic Tully-Fisher relation.
Recent observations from gas-rich galaxies match this relation well with the above value of $a_0$ \cite{mcgaugh}.
This undoubtedly suggests the importance of independent tests of MOND,
both in the realm of astrophysical observations and in the laboratory.

Equation~(\ref{mond1}) is consistent with at least two interpretations: the first, in which Newton's second law of motion is modified ({\it modified inertia})
\be
\vec F = m \vec a ~\mu \le( \frac{|\vec a|}{a_0} \ri) \equiv m \vec a_N~,
\la{nl1}
\ee
and the second, in which $\vec F = m \vec a$ remains intact, while only
for gravity the acceleration $a$ is given by the new law Eq.~(\ref{mond1})
({\it modified gravity}) \cite{milgrom2}.
Note that $m$ in Eq.(\ref{nl1}) is the {\it inertial} mass and that $\vec F$ can be gravitational {\it or} a
nongravitational force, and if the first interpretation is correct, then
deviations from Newtonian dynamics should be observed (in the MOND regime) for
any force, not necessarily gravitational.
In fact a couple of attempts to test the first interpretation with mechanical
oscillators 
did not report any deviations from Newton's second law down to accelerations of
$3\times10^{-11}~m/s^2$ \cite{vager} and $5\times 10^{-14}~m/s^2$, respectively \cite{gundlach},
i.e. up to $2000$ times lower than $a_0$,
from which one may be tempted to conclude that the MOND paradigm, or at least its first
interpretation is incorrect.
%
%
Note, however, that the large centripetal acceleration of Earth, far exceeding $a_0$
(except at the poles) is transmitted to the apparatus, making it virtually impossible to test the MOND paradigm
in conventional laboratories fixed rigidly to Earth, regardless of the sensitivity of the apparatus
\footnote{
Some authors have argued that MOND can perhaps be tested Earth at certain very precise locations and for very
short periods of time \cite{ignatiev}.
}.
%
%
However, Eq.(\ref{nl1}) may be tested, e.g. in spacecrafts, at a distance of
$0.1$ light years or more from our Sun, such that its acceleration is $a_0$ or less.
In addition, we note that the strong equivalence principle (SEP),
which asserts that all locally inertial frames are perfectly equivalent \cite{dewitt, will1993, weinberg}, 
 also guarantees testability of MOND in
experiments done in freely falling, i.e. locally inertial frames,
provided accelerations within the apparatus do not exceed $a_0$.
Although some observations related to open clusters and some theoretical formulations of MOND
suggest a potential violation of the SEP \cite{milgrom1} (manifesting in the form of the so-called
external field effect \cite{milgrom3}), these have not been verified by independent
observations, a fully satisfactory mechanism of these violations (including a covariant formalism)
remains to be understood, and quantitative predictions of anomalous accelerations within our solar
system due to the external field effect seem to depend on the interpretation, be sensitive to the $\mu$ function used,
and disagree with observations at least in some cases \cite{milgrom3}.
On the other hand, decades of careful testing have failed to detect any violation of the SEP
anywhere \cite{merkowitz}.
Accordingly in this paper,
we will not adhere to any specific theoretical formulation of MOND
and will not rule out the possibility that the SEP may still hold.
This is not inconsistent with any observation or fundamental theoretical principle to our knowledge.
We then propose tests of the MOND paradigm
for systems with internal accelerations in the regime
$a \leq a_0$
while the system as a whole is falling freely in a gravitational field with a
$g$ value that is much larger than $a_0$.
This effectively gets rid of Earth's acceleration.
%
The sensitivity levels achieved in experiments described in Ref.~\cite{gundlach}
will suffice for these experiments.
Examples of such freely falling laboratories include ``drop tubes,'' such as the
one in the NASA facility in Cleveland, Ohio; artificial satellites orbiting Earth, including, for example, the proposed Galileo Galilei satellite; and space-based missions orbiting the Sun and far away from Earth,
such as the Herschel Space Observatory~\cite{Pilbratt2010}\footnote{
We note that the proposed experimental tests of MOND with LISA pathfinder also assume a specific
formulation, namely, the nonrelativistic limit of the TeVeS theory
\cite{lisa}.}.
Possible logical outcomes include 
(i) failure to detect MOND, either supporting the conclusion of certain theories that
the {\it background acceleration}
$g\gg a_0$ renders a system Newtonian,
or that the first interpretation (if involving only nongravitational forces) or perhaps the
paradigm itself is wrong;
(ii) effects predicted by MOND are indeed detected, providing strong
and independent evidence in its favor as well as for the SEP;
and (iii) some deviations from Newtonian predictions are detected, which may be used to quantify the extent of
SEP violations. Later, we will show that such experiments may also
distinguish between the two interpretations of MOND. Thus, they seem to be something worth pursuing with technology already at hand.

\begin{figure}[h]
\begin{center}
\includegraphics[height=1.6in]{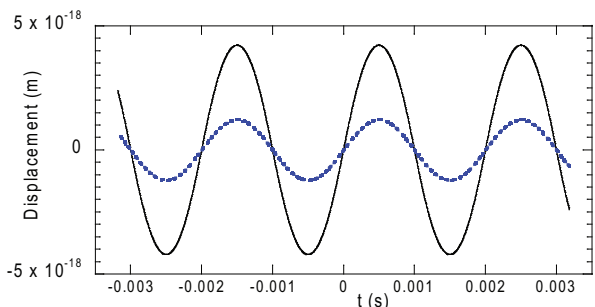}
\vs{-.5cm}
\caption{Steady-state response function $x(t)$ for deep MOND (solid) and Newtonian (dotted) with the same driving force.  A test mass of 16 kg is driven at 500 Hz with a force amplitude of $2\times 10^{-10}$ N.
\label{mondxt}}
\end{center}
\end{figure}

With this aim, we first consider Eq.~(\ref{nl1}), relating to the first MOND interpretation,
to a test mass $m$ suspended as a harmonic oscillator with spring constant
$k=m\omega_0^2$, subjected to a periodic force with angular frequency $\o \gg \o_0$ and displacement from equilibrium $x$, in a freely falling laboratory,
\be
m \frac{d^2x}{dt^2}~\mu\textstyle{ \le( \frac{1}{a_0}\frac{d^2x}{dt^2} \ri)} = F \sin(\o t) ~.
\la{hookemond}
\ee
Note that the forces involved are all nongravitational in nature, and
the displacement according to Newtonian mechanics ($\mu=1$) will be given by
\be
x_N(t) = - \frac{F}{m\o^2} \sin(\o t)~.
\ee
The above describes for example, the apparatus in Ref.~\cite{vager} near the sensitivity limit,
with a driving frequency in the range of several hundred Hertz.
Interestingly, it also describes modulated laser beam (with power $P_m$)
driven mirrors attached to the test masses in the LIGO gravitational wave detector, for which
$F=2P_m/c$ \cite{ligo2}
\footnote{We are assuming this auxiliary laser beam is almost parallel to the main beam.}
(the position is determined in both the above systems by interferometric methods).
With numbers close to the LIGO values, $P_m=30 mW$ and $m=16~kg$, the
Newtonian acceleration amplitude is $a_{N_{amp}}\equiv F/m=1.25\times 10^{-11}~m/s^2 \ll a_0$,
well within the MOND regime\footnote{The amplitude of motion $x_{N_{amp}}=a_{N_{amp}}/\o^2$ would equal
$1.3 \times 10^{-18}~m$ at a frequency of $\approx~500 Hz$, in agreement with the specification reported in Ref.~\cite{ligo2}.}.
Therefore, we write the MOND equation of motion derived from Eqs.~(\ref{nl1}) and (\ref{hookemond}):
\be
a = \frac{d^2 x}{dt^2} = \mbox{sgn}(\sin(\o t)) \sqrt{a_0 \frac{F}{m} |\sin(\o t)|}~,  \la{mondosc1}
\ee
where the sgn function takes care of the two signs of the acceleration,
and the absolute sign inside the square root follows from the definition of $\mu$.
Since the rhs of Eq.~(\ref{mondosc1}) is periodic with period $2\pi/\o$, it is most easily solved
by writing a Fourier series as follows: $ \mbox{sgn}(\sin(\o t)) \sqrt{|\sin(\o t)|}
= A_0/2 + \sum_{n=1}^\infty \le[ A_n \cos(n\omega t) + B_n\sin(n\omega t)\ri]$,
with $A_0=0$ since the average $ \langle\mbox{sgn}(\sin(\o t))\sqrt{|\sin(\o t)|} \rangle =0 $ over a period, and
$A_n=0~\forall n >0$ from the odd nature of the function.
Integrating twice gives
\be
x_M(t) = -\frac{\sqrt{a_0 F/m}}{\o^2}~\sum_{n=1}^\infty \le[ \frac{B_n}{n^2} \sin(n\omega t)\ri]~,
\ee
where the subscript $M$ signifies MOND.  In this case, the amplitude of displacement is given by $x_{M_{amp}}\approx \sqrt{a_0F/m}~B_1/\o^2$,
ignoring the $n^2$ suppressed higher harmonics
\footnote{This is further justified by the fact that all even-$n$ $B_n$ harmonics are zero, by symmetry considerations.}.
Note that the $1/\o^2$ scaling as in the Newtonian case is retained.
Thus, the ratio of the displacement amplitudes for the deep MOND to the Newtonian cases is
\be
\frac{x_{M_{(amp)}}}{x_{N_{(amp)}}} = \sqrt{\frac{a_0}{F/m}} B_1= \sqrt{\frac{a_0}{a_{N_{amp}}}}~B_1
\approx 3.1 B_1  \la{xmaxratio}
\ee
for the previously quoted parameter values.
A straightforward numerical computation gives $B_1=1.11, B_3=0.16~,B_5=0.07$, etc.
For this the amplitude for MOND is more than thrice the Newtonian value, as illustrated in Fig.~\ref{mondxt}.
Such a discrepancy would be easily detected by the LIGO interferometer, which can be linearly calibrated to within 2\%~\cite{ligo2}.
Also if the beam power, and hence the driving force, can be varied, then nonlinear behavior can also be verified if
the best fit curve of a plot of $\log(x_{M(amp)})$ vs $\log(F/m)$ is a straight line with slope
$+0.5$ in the deep MOND regime (the corresponding curve for the Newtonian should have slope $+1$).  Finally, Fourier analysis would show non-Newtonian harmonic peaks at odd-integer multiples of the fundamental.
\vs{-0.4cm}

Next we consider another highly sensitive tool historically used for testing fundamental physical principles, namely, the torsion balance \cite{Gundlach2009}, but once again operated in a free falling state.
Note that both gravitational and nongravitational forces are involved in this instrument.
In its simplest construction, two small spheres each of mass $m$ at the ends of a beam of length $L$, itself of negligible mass, are suspended from its center by a wire of torsion coefficient $\kappa$
(the latter involved a nongravitational force).
When a bigger mass $M$, is brought near each small mass, gravitational forces between the masses cause an angular displacement $\theta(t)$ from its equilibrium position $(\theta=0)$, which after several oscillations eventually settle down to a constant value $\theta=\theta_{s}$.\footnote{Some damping is assumed to be present.}
%
%
Then from Eqs.(\ref{mond1}) and (\ref{nl1}) (the first/modified inertia interpretation of MOND) and Hooke's law,
it follows that
\bea
\frac{GMmL}{r^2} -\kappa \theta &=& I~\frac{d^2\theta}{dt^2} \mu
\textstyle{\le( \frac{ \frac{L}{2} \frac{d^2\theta}{dt^2}}{a_0} \ri)}  \\
\&~~~  \theta_{s}  &=& \frac{GMmL}{\kappa r^2}~,
\eea
where $I=mL^2/2$ is the moment of inertia of the balance and $r$ is the center-to-center distance between small
and big mass at equilibrium.
While for the second/modified gravity interpretation in the deep MOND regime, one has
\bea
mL \sqrt{a_0 GM/r^2} -\kappa \theta &=& I~\frac{d^2\theta}{dt^2}  \\
\&~~~ \theta_{s}  &=& \frac{mL}{\kappa} \sqrt{a_0 GM/r^2}~.
\eea
Thus, if the first interpretation is correct, then the dynamics differs significantly from
Newtonian while the settling point remains the same, whereas for the second interpretation,
the dynamics remains Newtonian while the system settles down further away from the origin.
In either case, there is something new which can be potentially tested.
For example, in a $\log(\theta_{s})$ vs $\log(M)$ plot, a slope of $1$ would mean
the modified inertia interpretation (or no MOND at all), while a slope of $0.5$
would 
point toward MOND and its modified gravity interpretation.

 In a terrestrial laboratory,
a small platform freely falling in an evacuated chamber could serve as a sufficient locally inertial frame.  Small amounts of motion created intentionally to test MOND would be relative to this frame.  It is crucial though that there be no other sources of acceleration down to levels well below $a_0$.
For example, an object falling through residual gases at $10^{-8}$ Torr pressure will experience a velocity-squared type drag acceleration~\cite{vallado}, achieving levels near $a_0$ after a 10 m drop.  If absolute pressure in a vertical, tubular vacuum chamber is held under ultrahigh vacuum conditions, near $10^{-10}$~Torr, then drag effects will be negligible.

A drop-test experiment taking perhaps several seconds poses severe challenges for a torsion balance.  In the apparatus used by Gundlach {\it etal.}~\cite{gundlach}, the period of the balance was 795 s.  With a quality factor of 5000, there is simply not enough time for the instrument to settle down.  The weightless environment also causes unique challenges for the torsion balance.  Some of these challenges may be solved by using a second fiber located below the balance.  Alternatively, the larger masses and beam may be magnetically levitated, with a slight perturbation of a symmetric confining potential giving rise to a torsion coefficient~\cite{patitsas}. We note that for space-based tests of the equivalence principle, the torsion balance has been ruled out in favor of a concentric cylinders approach~\cite{Nobili2000}.

We note that the balance used in Ref.~\cite{gundlach} exceeded the sensitivity threshold for MOND tests by almost 4 orders of magnitude, which provides some room to trade off sensitivity in favor of 
exploring designs that might be more suitable for drop testing.  One important technical issue that must be raised is that the balance used in Ref.~\cite{gundlach} uses feedback to eliminate, as much as possible, the twist in the supporting fiber.  Testing MOND requires motion, so one should design an angle-of-deflection balance and avoid the use of (negative, proportional) feedback~\cite{Ritter1993}.  This will likely decrease the signal-to-noise ratio as well as exacerbate drift issues in the fiber.\footnote{Some of these issues apply to the LIGO interferometer mentioned above i.e. feedback on the test mass mirror would have to be eliminated in order to test MOND.}  The period of oscillation may also become lower~\cite{Ritter1993}.  Even so, we feel that there is enough room to implement a stiffer system (thicker wire) with a faster response time while still having the capability to test MOND.  Moreover, since a high-quality factor is not essential in testing MOND, as even just several oscillations as shown in Fig.~\ref{mondxt} would suffice, modest levels of damping could reduce the practical response time substantially.  One could also explore a sort of compromised feedback system which uses negative, proportional feedback with lower gain levels, thus allowing the fiber to twist somewhat and still keeping the beneficial aspects of the feedback, i.e. monitoring the less noisy feedback signal, reduced drift, and faster response time.  Even if the feedback cuts the actual motion down by $\approx$90\% the tradeoff may be worth it.     


For a spacecraft-based approach, we note that torques applied to the fiber due to the attraction of the test masses to the source masses would also act on the freely falling satellite.  A spacecraft design with source masses rigidly connected to a satellite frame designed with high cylindrical symmetry about the fiber axis would reduce this problem.

For terrestrial drop tests of MOND, one should look for a test-mass apparatus with a faster response time.  We propose a rather novel test-mass, located at the end of a micromachined cantilever, with 0.5~mm length and 2~$\mu$m thickness, serving as the oscillator fixed to the platform.  
We intend the driving force to be provided by a laser beam (as for the LIGO mirrors~\cite{ligo2}) also fixed to the platform and bouncing off a reflective surface on the cantilever. Calculations suggest a 25~mW laser beam modulated at $10$~Hz can exert enough force to cause cantilever acceleration levels of about $a_0$.
Displacement measurements can be made by reflecting another laser beam off the cantilever and using interference techniques, similar those used in atomic force microscopy.
We note that such cantilevers with optically reflective surfaces are available commercially.
The above setup would closely resemble that of Ref.~\cite{vager}, and should be classified as an interferometric apparatus, albeit with a novel test mass.  The proposed instrument should be compact and rigid in order to limit the detrimental effects of mechanical transients produced during the release process.  The release mechanism must also be carefully designed to minimize the creation of unwanted rotational motion, in particular with angular velocity components perpendicular to the plane of test-mass motion.  A good design would have the cantilever deflection in the horizontal plane.  We note that precision tests under terrestrial free fall conditions have been conducted to test the weak equivalence principle~\cite{faller1998}.  Our dropping mechanism could closely resemble the motorized platform used in these tests.  We note that this apparatus also has the advantage of being both simple and inexpensive, especially as compared to the LIGO apparatus. 

A microcantilever interferometer apparatus could also succeed if operated inside of an artificial satellite.  The use of high thermal conductivity alloys in combination with a compact design would also minimize the adverse effects of the radiometer effect due to temperature differences in the residual gas inside the spacecraft~\cite{Nobili2000}.  These differences are caused by the infrared radiation from Earth and can also be minimized by implementing multilayer passive thermal shielding.
For an atmospheric drag of about
$10^{-2}$~m/s$^2$ (typical for low Earth orbits of about $300$~km, and falling off exponentially~\cite{vallado}), the compact interferometer apparatus could succeed if placed inside of, and mechanically decoupled from the inside of, an orbiting spacecraft.  Drag levels below $a_0$ could be achieved if the residual gas pressure inside the spacecraft is $10^{-8}$~Torr~\cite{Nobili2000}.  The mechanical decoupling would have to be temporary, involving a repetitive release-and-catch mechanism.  The time available before requiring a catch process could be extended by use of ion thrusters to cancel the atmospheric drag ~\cite{Nobili2000}.  Thrusters could also compensate for the relative rotation of the stabilized spacecraft after a release.
The small physical dimensions of the cantilever results in a small cross section for cosmic ray events.  However, the small size means that events that do occur may be very disruptive~\cite{buchman1993}.  We would need to implement  pulse discrimination techniques to filter out any cosmic ray events.
Similarly, the small physical dimensions of the cantilever means tidal effects can be easily reduced to almost negligible values  of the order of
about $10^{-16}$~m/s$^2$ by aligning the platform and plane
of oscillation of the cantilever perpendicular to Earth's gravitational field~\cite{mtw}.
%
%
We emphasize that although a number of technical issues including the ones stated above
may need to be addressed in detail~\cite{dp2},
the two experiments we propose in freely falling frames appear to be within the reach of already existing technology or some adaptations thereof.
These would be capable of detecting
discrepancies from Newtonian dynamics, with one of them able to test the first interpretation, while the other would be capable of testing 
both interpretations. Thus, they deserve further study.


\vs{-.001cm}
We thank M. Milgrom and M. Walton for useful correspondence, the referees and editors for their useful comments and suggestions, as well as NSERC, Canada for support.

\end{document}